# Applications of CMOS technology at the ALICE experiment

**Domenico Colella** [a],[*] **for the ALICE Collaboration**

*University and INFN Bari,*
*Via E. Orabona 4, 170125, Bari, Italy*

*E-mail:* domenico.colella@ba.infn.it

Monolithic Active Pixel Sensors (MAPS) combine the sensing part and the front-end electronics in the same silicon layer, making use of CMOS technology. Profiting from the progresses of this commercial process, MAPS have been undergoing significant advances over the last decade in terms of integration densities, radiation hardness and readout speed. The first application of MAPS in high energy physics has been the PXL detector, installed in 2014 as the vertexer of the STAR experiment at BNL. In the same years, ALICE Collaboration started the development of a new MAPS with improved performances, to assemble a new detector to replace the Inner Tracking System used during LHC Run 1 and 2. This effort lead to the ALPIDE sensor, today successfully equipped in a large variety of systems. Starting from 2019, profiting from the experience acquired during the design of the ALPIDE sensor, the ALICE Collaboration embarked on a new development phase, the ITS3 project. Here the goal is to design the first truly cylindrical detector based on wafer-size sensors in 65 nm CMOS node. This new detector is expected to take data during LHC Run 4. ALICE Collaboration submitted a proposal for a new experiment, to be installed in place of the present detector system before the LHC Run 5. Building on the experience on MAPS acquired in the recent years, the idea is to design a compact all silicon detector, that will give unprecedented insight into the quark-gluon plasma characterization.

---

[*]Speaker

## 1. Introduction

ALICE (A Large Ion Collider Experiment) is a general-purpose heavy-ion experiment at the CERN LHC. During LHC Run 1 and 2 (2009-2018) the Collaboration has carried out a broad range of measurements to characterize the quark-gluon plasma and to study several other aspects of the strong interaction [1]. During LHC Run 3 and 4 (2022-2032) this study is focused on rare processes such as the production of heavy-flavor particles, quarkonium states, real and virtual photons and heavy nuclear states [2].

The most critical detector to achieve such a physics program is the closest one to the interaction point, the Inner Tracking System (ITS). The CMOS Monolithic Active Pixel Sensors (MAPS) technology has been chosen as the proper one to reach the required performances in terms of reduced material budget and improved readout speed and radiation hardness. The result of the intense R&D made by the ALICE Collaboration, is the ALPIDE sensor, representing a remarkable leap in terms of performances. The upgraded version of the ITS, the ITS2 [2], installed to take data during LHC Run 3 and 4, is the largest ever built MAPS based detector, covering a total surface of ~10 $m^2$.

Presently, the ALICE Collaboration is working on implementing further improvements in the usage of MAPS sensors for high energy physics applications. The ITS3 [3], a proposed upgrade of the ITS2, is being developed based on the observation that the 50 μm thin silicon tiles tend to bend naturally. The project is to assemble a detector almost self-standing in a real cylindrical shape, minimizing the material budget and improving the capability of the experiment to identify e.g. multi-charmed hadrons. The new detector will replace the three innermost layers of the ITS2 and be operated during LHC Run 4.

The success of the ITS2 assembly and the ongoing developments for the ITS3 detector, are guiding the development of the tracker for ALICE 3, the future experiment proposed by the ALICE Collaboration in 2022 [4]. The ALICE 3 detector concept is based on a compact layout within a superconducting magnet, with tracking and particle-identification subsystems mostly based on silicon sensors, that would extend the heavy-ion program at LHC during Run 5 and 6.

In the following three paragraphs, details about the applications of the CMOS technology within the ALICE experiment will be given.

## 2. The ALPIDE sensor

The demanding design requirements needed to pursue the physics program of the ALICE experiment in the LHC running conditions as expected for Run 3, required the development of a custom designed sensor, which involved a long R&D phase from 2012 to mid 2016. The ALPIDE sensor is fabricated in the 180 nm CMOS imaging sensor Tower Jazz process, which provides high resistivity (> 1 kΩ cm) p-type epitaxial layer on a p-type silicon wafer and deep p-wells, allowing for the implementation of full CMOS circuitry at the pixel level. The final design features a front-end circuitry with very low power consumption (<40 mW/$cm^2$), an integration time of ~2 μs and a pixel matrix readout approach based on a priority encoder. The chip measures 15 x 30 $mm^2$ and contains half million 27 x 29 µm$^2$ pixels, arranged in 512 rows and 1024 columns. Each pixel is equipped with amplification, discrimination and buffering, which allows a zero suppressed asynchronous readout. The most relevant features of ALPIDE are a spatial resolution of ~5 µm, an efficiency better than 99% and a fake hit rate <10$^{-10}$ hits/pixel/event. As can be seen in Figure 1, such a performance is verified in a rather large operation interval on irradiated sensors up to an ionizing dose (TID) of 500 krad and a non-ionizing energy loss (NIEL) of 1.7x10$^{13}$ 1 MeV n$_{eq}$/cm$^2$.

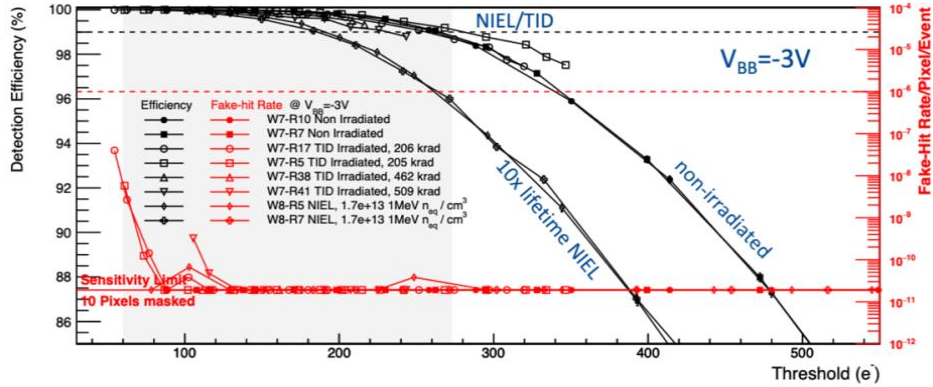

Figure 1. Detection efficiency (left axis) and fake hit rate (right axis) versus threshold of the ALPIDE chip irradiated with different NIEL and TID doses. [2]

A total of 24k ALPIDE sensors have been integrated in staves to form the seven layers of the ITS2 [2]. The three innermost layers, with radii spanning from 22 mm to 39 mm, are equipped with 50 µm thick ALPIDEs, while the four outermost layers, from 196 mm to 393 mm, are equipped with 100 µm thick ALPIDEs, profiting from the higher chip production yield to cover the larger surface.

A forward high position resolution detector for muon tracking, the Muon Forward Tracker (MFT) [2], also equipped with ALPIDE sensors, has been installed as well to operate during LHC Run 3 and 4. It has a projective geometry based on five disks, coaxial with the beam pipe, with disk radius increasing with the distance to the interaction point. Its primary goal is to improve the pointing resolution of muons; the reconstruction procedure foresees the matching of the tracks provided by a gaseous detector downstream of an hadron absorber to those obtained from the MFT upstream the absorber.

The ALPIDE sensor was also chosen to equip an highly granular electromagnetic calorimeter as part of a new detector, the Forward Calorimeter (FoCal) [6], to be installed in the ALICE experiment and operated during LHC Run 4. The FoCal will carry out a rich forward physics program, including prompt and isolated photons, identified π$^0$ and other neutral mesons, J/$\psi$ and its excited states.

### 3. The ITS3 upgrade

R&D targeting a further improvement in tracking and vertexing capabilities, especially at transverse momentum below 1 GeV/c, started within the ALICE

Collaboration to develop the first fully cylindrical detector based on wafer-size MAPS sensors, the ITS3 [3]. The concept of this detector is based on the following R&D directions:
- explore the capability of thin silicon (< 50 μm) to be bent and profit from the stiffness deriving from bending to remove most of the mechanical support structures;
- to cover the needed acceptance of one half-layer with a single sensor and consequently decrease the segmentation and reduce the mechanical supports, profit from the 300 mm wafer, used by the Tower Partner Semiconductor to produce sensors in 65 nm CMOS node;
- design the first MAPS used in high energy physics at 65 nm CMOS node, implementing also the stitching technique to actually obtain a sensor of maximum sizes 10 cm x 27 cm;
- profit from the low power consumption for this CMOS technology to try to cool it simply using an air flow, consequently removing all the related material budget from the sensible area.

Unaffected performances of MAPS when bent at the target radius of the innermost layer of the ITS3 (~18 mm), have been measured using ALPIDE sensors and test structures in 65 nm [3]. The latter, firstly received in 2021, undergone a large characterization campaign that completely validated the technology for the ALICE application [7]. As can be seen in Figure 2, the performances in terms of radiation hardness extend by at least, one order of magnitude with respect to the ALPIDE sensors. This has been possible by implementing process optimizations, explored as follow up of the ALPIDE development effort to reach the full depletion in the epitaxial layer, improving the timing performance and the radiation tolerance [6]. In the following step, the first large-area sensor implementing the stitching technique has been designed. This technique allows to extend the size of the mask that is used in photo-lithography process, connecting independent reticles already at wafer production stage. The first prototypes have been received in summer 2023 and immediately characterized in terms of detector efficiency and production yield [8].

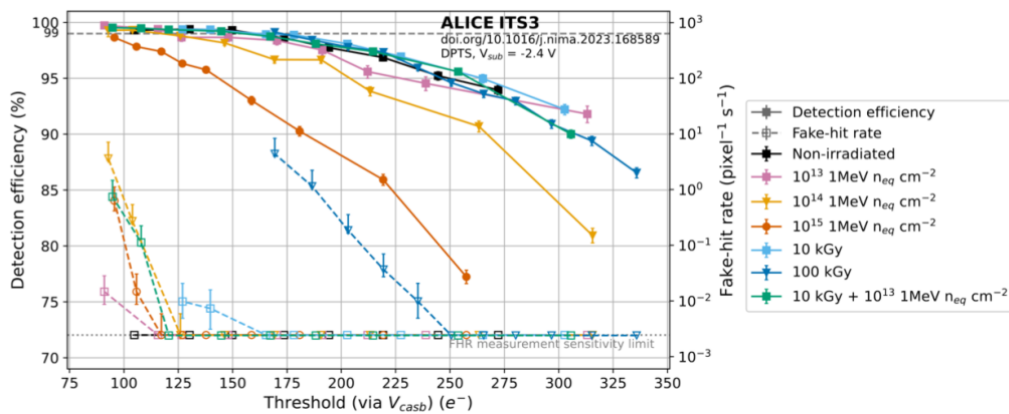

Figure 2. Detection efficiency (left axis) and fake hit rate (right axis) versus threshold for the first digital matrix in 65 nm CMOS irradiated with different NIEL and TID doses. [3]

Large dimension silicon sensors are bent in cylindrical shape by means of a mandrel and a tensioned mylar foil. Carbon foam spacers, 6 mm thick, separate the sensor layers and the shape is granted through an external carbon fiber cylindrical shell. Air is blown between the layers through ducts at a velocity of approximately 8 m/s, granting uniformity along the sensor and an increase to the inlet temperature of maximum 5 °C [9].

### 4. The ALICE 3 tracker

Despite the expected progress in understanding the condensed matter of QCD, several key measurements will still be missing after LHC Run 3 and 4. The ALICE Collaboration is developing a new detector, to be operated during LHC Run 5 and 6, whose key requirements are: a retractable vertex detector with an unprecedent pointing resolution, a compact and lightweight all-silicon tracker combined with a superconducting magnet system and extensive particle identifications with different dedicated subsystems [4]. The MAPS will be the largely used technology in the experiment, profiting from the development ongoing within the ITS3 and requiring new step forward on many aspects.

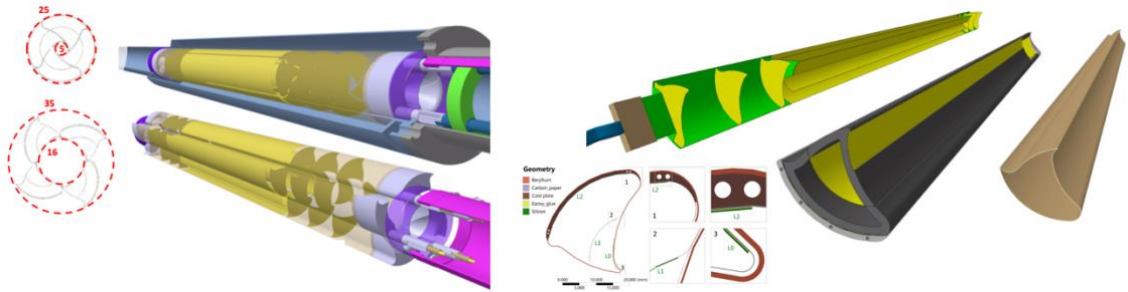

Figure 3. IRIS detector retractable capability (right) and petal module (left).

The innermost detector, dedicated to vertexing, will consist of three layers of wafer-size, ultra-thin, curved MAPS inserted inside the beam pipe (Figure 3). The modules are contained inside independent secondary vacuum cases and, through a remote control system, will be retractable allowing to bring, at stable beams, the first layer to 5 mm from the collision point. The main R&D challenges concern mechanical support, cooling, material degassing and radiation tolerance of the sensors (300 Mrad + $1\times10^{16}$ 1 MeV $n_{eq}/cm^2$).

The vertex detector is complemented by 8 cylindrical layers, external to the beam pipe and with radii up to 80 cm, and 9 forward disks, covering a total surface of ~70 $m^2$. Here the performance requirements are less stringent, and the layers can be equipped by more traditional staves, conceptually like the ones of ITS2. Due to the large surface to be covered, the assembly and testing procedures of the modules must be automatized as much as possible. Exploration of interest and availability with external companies is ongoing.

Finally, MAPS are also considered as possible technology for the development of a Time of Flight (TOF) detector for the particle identification, in parallel with SiPM and LGAD. The main requirement for the detector is a time resolution of ~20 ps. The ARCADIA R&D project by INFN, is developing a fully depleted MAPS in a 110 nm CMOS process at LFoundry, adding a gain layer below the pixel pad. First prototype sensor, the MadPix, has been characterized in test beam campaign in October 2023 and a new version will be available in 2024 with improved front-end electronics, characterized by a lower jitter, to achieve the target 20 ps total time resolution.

## 5. Summary

The ALICE experiment demonstrated with ITS2 detector that MAPS technology is mature enough to equip large and critical detectors in high energy physics experiments. The sensor R&D carried on toward the full depletion of the epitaxial layer, for better timing performance and radiation hardness, is pushing the performance of these devices. As well, the introduction of the stitching technique is opening new scenarios for low material budget layouts for MAPS based detectors. Finally, the ambitious design of the ALICE 3 tracker is pushing further the technical advancements in the application of MAPS, targeting sensors with even higher radiation hardness and better timing resolution.